\documentclass[12pt,preprint]{aastex}

\usepackage{longtable}
\usepackage{multirow}
\usepackage{url}
\usepackage{graphicx}
\usepackage{lipsum}

\usepackage{natbib}
\usepackage[utf8]{inputenc}
\usepackage{amsmath}

\usepackage{xspace}
\usepackage{lscape}
\usepackage{color}

\begin{document}
\title{Investigating the effect of cosmic opacity on standard candles}
\author{J. Hu\altaffilmark{1}, H. Yu\altaffilmark{1} and F. Y. Wang\altaffilmark{1,2}{*}}
%\author{J. Hu$^{1}$}
%\author{H. Yu$^{1}$}
%\author{F. Y.Wang$^{1,2}$} \email{fayinwang@nju.edu.cn}
\affil{
$^1$ School of Astronomy and Space Science, Nanjing University, Nanjing 210093, China\\
$^2$ Key Laboratory of Modern Astronomy and Astrophysics (Nanjing University), Ministry of Education, Nanjing 210093, China\\
} \email{fayinwang@nju.edu.cn}

\begin{abstract}
Standard candles can probe the evolution of dark energy in a large
redshift range. But the cosmic opacity can degrade the quality of
standard candles. In this paper, we use the latest observations,
including type Ia supernovae (SNe Ia) from JLA sample and Hubble
parameters, to probe the opacity of the universe. In order to avoid
the cosmological dependence of SNe Ia luminosity distances, a joint
fitting of the SNe Ia light-curve parameters, cosmological
parameters and opacity is used. In order to explore the cosmic
opacity at high redshifts, the latest gamma-ray bursts (GRBs) are
used. At high redshifts, cosmic reionization process is considered.
We find that the sample supports an almost transparent universe for
flat $\Lambda$CDM and XCDM models. Meanwhile, free electrons deplete
photons from standard candles through the (inverse) Compton
scattering, known as an important component of opacity. This Compton
dimming may paly an important role in future supernova surveys. From
analysis, we find that about a few percent cosmic opacity is caused
by Compton dimming in the two models, which can be correctable.
\end{abstract}

\keywords{cosmology: theory - distance scale}

\maketitle

\section{Introduction}\label{sec:introduction}
In 1998, the accelerating expansion of the universe was discovered
by measuring the relation between redshift and distance of SNe Ia
\citep{Riess98,Perlmutter99}. The physical origin of accelerating is
still debated. The term ``dark energy'' is put forward to explain
the accelerating universe. Meanwhile, the modification of equations
governing gravity can also explain the acceleration of the universe
\citep[i.e.,][]{Capozziello2002}. Besides SNe Ia, other
observations, such as cosmic microwave background (CMB)
\citep[i.e.,][]{SDN01}, baryonic acoustic oscillations (BAO)
\citep[i.e.,][]{ED01}, Hubble parameters \citep[i.e.,][]{JR01}, and
gamma-ray bursts (GRBs) \citep[i.e.,][]{Wang15}, can probe the
nature of accelerating expansion.

SNe Ia are ideal standard candles to probe dark energy. But several
effects can degrade their quality, such as the dust in light path
\citep{Av}, the possible intrinsic evolution in SN luminosity,
gravitational lensing magnification \citep{Holz98}, peculiar
velocity \citep{HG06}, and so on. These processes will degrade the
standard candle usefulness of SNe Ia. Besides the above effects,
Compton dimming due to free electrons deplete photons from standard
candles by the (inverse) Compton scattering can cause systematic
error for cosmological studies \citep{PZ}. These effects can degrade
the evidence of accelerating expansion, or even mimic the dark
energy behavior. So comprehensive study of the cosmic opacity is
needed. Especially for the Wide Field Infrared Survey Telescope
(WFIRST) era, which can detect more than 2000 SNe Ia \citep{GJ01}.
If the cosmic opacity is not corrected, it will not increase
statistical errors, but may also systematically bias the
cosmological parameters.

Over the past several years, the cosmic distance duality (CDD)
relation has been widely used to test the systematic errors and
opacity in SNe Ia observations. The CDD relation reads
\citep{eth,
ellis07}
\begin{equation}
\frac{{{D_L}}}{{{D_A}}}{(1 + z)^{ - 2}} = 1,
\end{equation}
where $D_L$ is the luminosity distance, and $D_A$ is the angular
diameter distance. We must note that the cosmic opacity has no
effect on the angular diameter $D_A$ \citep{weinberg}. It is valid
for all cosmological models based on Riemannian geometry. The bases
of this relation are that the number of photons is conservative and
the photons travel along the null geodesics in a Riemannian
spacetime \citep{ellis07}. But the conservation of photons may be
violated in a wide range of well-motivated models. In modern
astronomy, the CDD relation plays a significant role. In order to
test this relation, many works have been performed. For example,
\cite{BK04} found a 2$\sigma$ violation of CDD relation using $D_L$
from SNe Ia and $D_A$ from FRIIb radio galaxies. The angular
diameter from X-ray observations of galaxy clusters also has been
used to probe the CDD relation \citep{Holanda6}. Similar works have
also been done by other authors \citep{mx,rr}. \cite{Rasanen16} used
CMB anisotropies to test the CDD relation. This relation is also
applied extensively. \cite{xx} and \cite{Cao2016} used the CDD
relation to test the gas mass density profile of galaxy clusters.
\cite{Evslin2016} calibrated the distances of SNe Ia using the CDD
relation.

A powerful method to study the opacity of the universe is using the
standard candles to detect possible CDD deviations, such as SNe Ia
and GRBs. For example, \cite{Av2} adopted a modified CDD relation
\begin{equation}
D_L = D_A(1 + z)^{2 + \varepsilon }
\end{equation}
to constrain the cosmic opacity by combining the SNe Ia data
\citep{ko} with the measurements of the Hubble expansion over
redshift range $0 <z <2 $ \citep{DS01}. In the flat $\Lambda$CDM
model, they found $\varepsilon = - 0.04_{ - 0.07}^{ + 0.08}$
$(2\sigma)$. In \cite{Av}, they marginalized over the parameter
$H_0$ and used SNe Ia alone to constrain parameters $\Omega_m$ and
$\varepsilon$. \cite{LP} presented some tests for the cosmic opacity
with observational data including the Union 2.1 SNe Ia sample and
galaxy cluster samples compiled by \cite{Fil} and \cite{bo}. They
found that an almost transparent universe is favored by the sample
\citep{LP}. Basing on the validity of the Amati relation,
\cite{Holanda2} determined the cosmic opacity at high redshifts
using GRBs, and found that a transparent universe is favored. Strong
gravitational lensing systems are also used to probe the CDD
realtion \citep{Lia,Holanda5}.

Compared with previous papers, our paper has three advancements to
this field. First, it must be noted that previous studies directly
used the luminosity distances of SNe Ia, which are derived in the
concordance cosmology \citep[i.e.,][]{Av2,LP}. The luminosity
distances depend on the light-curve fitting parameters and
cosmological models \citep{ko,MR01}. So the derived results are
biased by the assumed cosmological model. Here, in order to avoid
this problem, we perform a global fitting for the SNe Ia light-curve
parameters, cosmological parameters and cosmic opacity. Second, we
also investigate the cosmic opacity at high redshifts, where the
fraction of electrons is evolving with redshift. The reionization
process is considered. The cross section of Compton scattering for
high-energy photons is also a function of redshfit. Third, the
contribution from Compton scattering effect to the cosmic opacity is
constrained for the first time. In this paper, we investigate the
cosmic opacity with SNe Ia, long GRBs and Hubble parameter data. We
pay special attention to the Compton scattering effect. This paper
is organized as follows. In next section, we describe the cosmic
opacity and Compton scatter extinction. In section 3, the
observational data used in the statistical analysis are presented.
The corresponding constraints on the cosmic opacity are given in
section 4. The paper is finished with a summary of the main results
in the conclusion section.

\section{Cosmic opacity and Compton scattering}
Since photons can be scattered with free electrons and interstellar
medium when travel from the source to the observer, the received
photons number will be reduced. The distance modulus derived from
standard candles will increase the systematic error. Any process
reducing photon number would increase the luminosity distance of the
source and dim its luminosity. Following \cite{Av}, we regard
$\tau(z)$ as the opacity from the $z=0$ to the resource redshift due
to extinction. Then, the received flux will decrease with a factor
$e^{-\tau(z)}$. So the relation between observed luminosity distance
$D_{L,obs}$ and theoretic luminosity distance $D_{L,th}$ is
\begin{equation}
 D_{L,obs} = D_{L,th}e^{\frac{\tau(z)}{2}}.
\end{equation}
The observed distance modulus is given by
\begin{equation}
 \mu_{obs}(z)=\mu_{th}(z)+2.5(\log_{10}(e))\tau(z).
\end{equation}
For flat FLRW cosmology, the distance modulus is
\begin{equation}
D_{L,th}(z)=(1+z)\frac{c}{H_0}\int^{z}_{0}\frac{dz}{E(z)},
\end{equation}
and
\begin{equation}
E(z)=H(z)/H_0=\sqrt {\Omega _m(1 + z)^3 + (1 - \Omega
_m)(1+z)^{3+3w}}.
\end{equation}
Combining equations (2) and (3), we obtain the exactly form of
cosmic opacity
\begin{equation}
\tau=2\varepsilon\ln(1+z).
\end{equation}

\subsection{The optical depth of Compton scattering}
Compton scattering is the inelastic scattering of the photon by a
charged free electron. The optical depth for Compton scattering is
\begin{equation}
\tau_c(z)=\int\sigma_Tn_e(z)dl\\
=-(1+y)\sigma_Tc\int^{z}_{0}n_H(z)Q_{H_{II}}(z)\frac{dt}{dz}dz,
\end{equation}
%\begin{equation}
%n^{free}_e(z)=n_H(z)X_e(z)
%\end{equation}
where ${\sigma_T}$ is the Thomson cross section, $n_e$ is the free
electron density, $c$ is the light speed, and $z$ is the redshift.
In the above equation, $n_H(z)=1.905\times10^{-7}(1+z)^3$ cm$^{-3}$
is the hydrogen number density at redshift $z$, and $y$ is a factor
which is introduced by including the ionization of helium. Because
the reionization epoch contains both hydrogen and helium. The mass
fractions of hydrogen and helium are $X=1-Y$ and $Y=0.24668$
\citep{arxiv1}, respectively. We assume that the helium was only
ionized once. So we derive $y=Y/(4X)\approx0.082$. $Q_{H_{II}}(z)$
is defined as the volume filling fraction of ionized hydrogen, which
can be calculated from the differential equation
\citep{mad99,balo01,wfy}
\begin{equation}
\dot{Q}_{\rm H_{\rm II}}={\dot{n}_{\gamma}(z) \over (1+y)n_{\rm
H}(z)}-\alpha_{B}C(z) (1+y)n_{\rm H}(z) Q_{\rm H_{\rm
II}}\label{reionize}.
\end{equation}
In this equation, $\alpha_B=2.6\times10^{-13}$ cm$^{-3}$s$^{-1}$ is
the recombination coefficient for electron with temperature at about
$10^4$ K. $\dot{n}_{\gamma}(z)$ is the rate of ionizing photons
escaping from the stars into the IGM, which can be derived from
\begin{equation}
\dot{n}_\gamma(z)=(1+z)^3\frac{\dot{\rho}_{\ast(z)}}{m_B}N_\gamma f_{esc},
\end{equation}
where $(1+z)^3$ is used for converting the comoving density into the
proper density, $\dot{\rho}_{\ast}(z)$ is the star formation rate
(SFR), $m_{\rm B}$ is the baryon mass, $N_\gamma$ are the number of
ionizing UV photons released per baryon, and $f_{\rm esc}$ is the
escape fraction of these photons from stars into IGM. The escape
fraction is not well constrained from observations.
$f_{esc}\leqslant0.2$ is the average value suggested by \cite{mao07}
and \cite{robe15}. Other similar value are reported. For instance,
\cite{raso06} found that $f_{esc}$ evolves from $\sim 1 - 2$ percent
at $z = 2.39$ to $\sim$ 6 \text{-} 10 percent at $z = 3.6$ from star
forming regions in young galaxies. \cite{Hay11} proposed a redshift
evolution of $f_{esc}$. In this work, we take the value of
$N_\gamma$ as $\sim$ 4000 and the escape fraction $f_{esc}\simeq
0.1$. $C\equiv\langle n^2_{H_{II}}\rangle/{\langle
n_{H_{II}}\rangle}^2$ is the clumping factor of the ionized gas. Its
value decreases with increasing redshifts from some numerical
simulations \citep{gne97,shu12} and semi-analytical studies
\citep{mad99,chos00}. Following \cite{shu12}, we take
\begin{equation}
C(z) = \left\lbrace
      \begin{array}{ll}
        2.9 &\quad \text{if}\quad  z <5,\\
        2.9(\frac{1+z}{6})^{-1.1}, & \quad \text{if}\quad  z \geq 5.
      \end{array}
    \right.
\end{equation}
$\dot{\rho}_{\ast(z)}$ is the SFR. The SFR derived by \cite{wfy} is
used. Then we can solve the differential equation (9) to obtain
$Q_{H_{II}}$. The result is shown in figure 1.

\subsection{The Compton scattering optical depth for SNe Ia}
Following \cite{hu95} and \cite{balo01}, the equation (8) with a
constant ionization fraction can be expressed as
\begin{equation}
\tau_c(z)=0.0461(1+y)Q_{H_{II}}(1-Y_p)\frac{\Omega_bh}{\Omega_m}\{[1-\Omega_m+\Omega_m(1+z)^3]^{\frac{1}{2}}-1\},
\end{equation}
in the flat $\Lambda$CDM model by neglecting the radiation term. At
redshift range $0<z<3$, a constant ionization fraction $X_e(z)=1$ is
adopted, which is reasonable for SNe Ia. The optical depth can
increase the distance modulus with a relation $\Delta\mu=1.086\tau$
from equation (4). Figure 2 shows the Compton scattering effect on
the distance modulus. From this figure, we can see that the value of
$\Delta\mu$ is increasing with redshift, and the Compton scattering
dims the supernova flux by 0.003 mag at $z = 1$ and 0.01 mag at $z =
2.35$, respectively. This dimming is too faint to rule out the
existence of dark energy. However, its effect can not be negligible
for future SNe Ia surveys plan such as WFIRST, which will measure
$\sim$ 2700 SNe Ia to $z \sim 1.7$. For future surveys, the major
statistical uncertainty is the SN intrinsic fluctuations. With the
SNe Ia number $N$, the intrinsic fluctuations are reduced to a level
of ${\sigma _\mu }{\rm{/}}\sqrt N$ mag, where ${\sigma _\mu }$ is
the intrinsic dispersion in SN luminosity. It means that the Compton
dimming effect must be corrected. Otherwise the induced systematic
errors would be comparable to the statistical errors. From above
analysis, we conclude that the Compton scattering can be
correctable, as discussed by \cite{PZ}.

\subsection{The Compton scattering dimming for GRBs}
The photons emitted from GRBs are  different from those from SNe Ia.
First, the energy of the GRB photons are much more energetic. At
high energies, cross section of the Compton scattering is
suppressed. So more photons can escape from scattering to the
observer. Second, GRBs can be observed at high redshifts.
High-energy photons have much more probability to interact with free
electrons. The optical depth of Compton scattering for high-energy
photons can be written as
\begin{equation}
\tau_c(z)=-(1+y)c\int^{z}_{0}\sigma(x)n_H(z)Q_{H_{II}}(z)\frac{dt}{dz}dz,
\end{equation}
where $\sigma (x) = \sigma({E_0}(1 + z)/{m_e}{c^2})$
is given by the Klein-Nishina formula \citep{Rybicki76}
\begin{equation}
\sigma (x){\rm{ = }}\frac{3}{4}{\sigma _T}[(1 + x)\frac{{2x(1 + x)/(1 + x) - \ln (1 + 2x)}}
{{{{\rm{x}}^3}}} + \frac{{\ln (1 + 2x)}}{{2x}} - \frac{{1 + 3x}}{{{{(1 + 2x)}^2}}}].
\end{equation}
Here, $E_0$ is the observed energy of $\gamma-$ray photons. The
future SOVM (Space-based multiband astronomical Variable Objects
Monitor) mission, will detect some GRBs at $z>10$ \citep{wei16}. At
these high-redshifts, the hydrogen is not completely ionized. The
parameter $Q_{H_{II}}$ is a constant in equation (13), and the
reionization process must be considered. We use the reionization
process described in Section 2.1 to calculate the optical depth. The
systematic shift in distance modulus $\Delta\mu$ due to Compton
scattering is shown in figure 3. It is obvious that the effect of
Compton scattering for low-energy photons is significant, because
the cross section is suppressed for high-energy photons. The
evolution of $\Delta\mu$ becomes flat at high redshifts, due to few
free electrons from reionization. The $\Delta\mu$ caused by Compton
dimming increases with reshift. Its value can reach to 0.01-0.04
mag, which is smaller than the intrinsic error of GRB distance
\citep{JW}. So we can ignore it if the number of GRBs is less than
100 and the redshift of GRB is not very high. However, if more than
100 high-redshift long GRBs will be used to study cosmology, the
Compton dimming is non-negligible.

\section{Data set}
In this section, we will show the data sets. These data sets will be
used to constrain the cosmic opacity and cosmological parameters.
Unlike previous works, we try to global fit the SNe Ia light-curve
parameters, cosmological parameters and the cosmic opacity.

\subsection{SNe Ia sample}
In this work, we use 740 SNe Ia from the ``joint light-curve
analysis" sample compiled by \cite{MR01}. The redshift range is from
$0.01$ to $1.299$.  This sample includes SNe Ia from different
surveys.In their work, they regard the possible extinction as
systematic uncertainty. In order to avoid the effect of cosmological
model effect, the parameters of the SNe Ia light-curve, cosmological
parameters and the cosmic opacity are fitting simultaneously.
Therefore, only the statistical error which from error propagation
of light-curve fitting uncertainties and the variation of magnitudes
caused by the intrinsic variation in SN magnitude are needed to
consider in our work. The possible extinctions are all regarded as
cosmic opacity. The distance modulus is written as
\begin{equation}
\mu  = m_B^ \star  - ({M_B} - \alpha  \times {X_1} + \beta  \times C),
\end{equation}
where $m_B^ \star$ is the observed peak magnitude in rest-frame $B$
band. $\alpha$ and $\beta$ are nuisance parameters which describe
the stretch-luminosity and color-luminosity relations, reflecting
the well-known broader-brighter and bluer-brighter relations,
respectively. The nuisance parameter ${M_B}$ represents the absolute
magnitude of a fiducial SNe Ia and is found to depend on the
properties of host galaxies, e.g., the host stellar mass
($M_{stellar}$). Here, we follow the procedure in \cite{Conley11} to
approximately correct for this effect by a simple step function:
\begin{equation}
  M_B = \left\lbrace
      \begin{array}{ll}
        M^1_B &\quad \text{if}\quad  M_\text{stellar} < 10^{10}~M_{\odot}\,,\\
        M^1_B + \Delta_M & \quad \text{otherwise.}
      \end{array}
    \right.
\end{equation}

\subsection{GRB sample}
For GRBs, we use the GRB data given in \cite{JW}. They use the
$E_{iso}$-$E_p$ correlation \cite{Amati02} to build the Hubble
diagram. \cite{JW} combine their 42 GRBs and 109 GRBs from
\cite{Amati08} and \cite{Amati09}. The $E_{iso}$-$E_p$ correlation
can be written as
\begin{equation}
\log \frac{{{E_{iso}}}}{{{\rm{erg}}}} = c + d\log \frac{{{E_p}}}{{{\rm{kev}}}},
\end{equation}
where parameters $c$ and $d$ are free parameters, $E_{iso}$ is the
isotropic equivalent energy, and $E_{p}$ is the peak energy of $\nu
F_{\nu}$ spectrum, which has been corrected into the cosmological
rest frame. In their work, they calibrate 90 high-redshift GRBs in
the redshift range from 1.44 to 8.1 with a fixed value of $H_0$. We
constrain the cosmological parameters and the cosmic opacity use
this sub-sample \citep{JW}. In order to consider the effect of
Compton dimming, we show the value of $\Delta\mu$ for this sample as
dots in figure 3, which is derived from equations (4) and (13). The
error bar is due to the uncertainty of the observational peak energy
of GRBs. The value of $\Delta\mu$ caused by Compton dimming is far
less than the top black dash line. Therefore, we can ignore this
effect in following work.

\subsection{$H(z)$ sample }
The 19 Hubble parameter data given in \cite{J01}, \cite{DS01}, and
\cite{MM011} are used in this work. The redshift range of these
Hubble parameters is from 0.10 to 1.75. Because $H_0$ will affect
the final results, we regard $H_0$ as a free parameter.

\section{Results}
The maximum likelihood analysis is used to constrain the parameters.
The $\chi^{2}$ fitting expression is
\begin{equation}
\chi^2=\sum\limits_i^n\frac{[\mu_{obs}-\mu_{th}(z_i)-1.086{\tau
(z_i)}]}{\sigma^2_{\mu}}+\chi^2_{H(z)}
\end{equation}
In our analysis, we adopt the cosmic opacity from equation (7). The
parameter  $\varepsilon$ is regarded as a constant. For data of SNe
Ia and $H(z)$, The $\mu_{obs}$ for SNe Ia is written as equation
(15). $\sigma^{2} _{\mu}=\sigma^2_{\mu,stat}+\sigma^2_{\mu,sys}$ is
the distance modulus uncertainty. $\sigma^2_{\mu,stat}$ is the
propagated error from the covariance matrix of the light-curve
fitting, and $\sigma _{\mu,sys}$ is the systematic error due to the
intrinsic variation in SNe Ia magnitude. The value $\sigma
_{\mu,sys}$ is calculated in \cite{MR01}, which is not depend on a
specific choice of cosmological model. $\mu_{th}$ is the theoretic
distance modulus which is depend on cosmological model. $\chi^2
_{H(z)}$ is the $\chi^{2}$ fitting of Hubble parameter data, which
can be calculated by
\begin{equation}
\chi^2 _{H(z)} = \sum\limits_i^m
{\frac{[H_{obs}(z_i)-H_{th}(z_i)]^2}{\sigma _{H_{obs}}^2}},
\end{equation}
where the $H_{obs}$ is the observation value, $H_{th}$ is the
theoretic Hubble expansion rate related to cosmological model, and
$\sigma _{H_{obs}}$ is the error of $H_{obs}$. For the GRB data,
because \cite{JW} calibrated the distance moduli by fixing
$H_0=67.8$ km~s$^{-1}$~Mpc$^{-1}$, so the value of $H_0$ is fixed
when using the GRB data. We use the Markov chain Monte Carlo (MCMC)
method to fit the parameters of the SNe Ia light-curve, cosmological
parameters and the cosmic opacity simultaneously. Our program is
based on the public emcee Python module \citep{fomc12}. The
algorithm of emcee has several advantages over traditional MCMC
methods and it has excellent performance as measured by the
autocorrelation time.

\subsection{Flat $\Lambda$CDM}
In this model, the equation of state $w$ in equation (6) has a fixed
value with $w=-1$. When using the SNe Ia + $H(z)$ data, the free
parameters are $M_B$, $\alpha$, $\beta$, $\Delta M$, $H_0$,
$\Omega_m$, and $\varepsilon$. We use the emcee Python module to fit
these parameters simultaneously. The fitting result is shown in
figure 4. The 2-D regions and 1-D marginalized distributions with
1$\sigma$ and 2$\sigma$ contours for the parameters $M_B$, $\alpha$,
$\beta$, $\Delta M$, $H_0$, $\Omega_m$, and $\varepsilon$ are shown.
The fitting results of parameters are presented in table I. The
value of $\varepsilon$ is $0.0226^{+0.0403}_{-0.0451}$, which
indicates an almost transparent universe. For GRB+$H(z)$ data, there
are only two free parameters: $\Omega_m$ and $\varepsilon$. The
fitting results are shown in figure 5 and table I. The value
$\varepsilon=0.00718^{+0.0486}_{-0.0492}$ also supports a
transparent universe.

\subsection{Flat XCDM}
In a flat XCDM cosmology, the parameter $w$ in equation(6) is a free
parameter. When using the SNe Ia + $H(z)$ data, the free parameters
are $M_B$, $\alpha$, $\beta$, $\Delta M$, $H_0$, $\Omega_m$, $w$ and
$\varepsilon$. Using the same method as above, we can perform the
simultaneously fitting of these parameters. The 2-D regions and 1-D
marginalized distributions with 1$\sigma$ and 2$\sigma$ contours for
the parameters $M_B$, $\alpha$, $\beta$, $\Delta M$, $H_0$,
$\Omega_m$, and $\varepsilon$ are shown in figure 6 and table I.
They are shown in the fourth and the last column of table I. The
value of $\varepsilon=$ is $0.0517^{+0.0617}_{-0.0659}$. For
GRB+$H(z)$ data, there are three parameters: $\Omega_m$, $w$ and
$\varepsilon$. The fitting results are shown in figure 7 and table
I. The value $\varepsilon=0.0718^{+0.0497}_{-0.0491}$ also indicates
a transparent universe.

\subsection{Considering the effect of Compton dimming}
Because the effect of Compton dimming can be estimated, the residual
opacity can be derived. We try to eliminate the known opacity due to
Compton scattering, and explore the contribution by unknown part. In
the equation (12), we get the optical depth of Compton scattering of
SNe Ia. After subtracting the optical depth of Compton scattering
from total cosmic opacity, we repeat the above analysis to obtain
the residual opacity $\tau_r$. In flat $\Lambda$CDM model, the
results from SNe Ia + $H(z)$ are shown in figure 8, which gives
$\varepsilon=0.0212^{+0.0382}_{-0.0413}$. Constraints on parameters
are shown in the column 3 of table I. Similar results are also shown
in figure 8 and column 5 of table I for XCDM model. Comparing the
second and third columns in table I, it can be seen that the effect
of Compton scattering can cause about 5\% cosmic opacity in
$\Lambda$CDM model. For XCDM model, a similar percentage is found.
So Compton scattering can contribute about a few percent of cosmic
opacity. It's obvious that the sample supports an almost transparent
universe for both cosmological models.

\section{Conclusions and discussion}

In this paper, we use the latest observations, including SNe Ia from
JLA sample and Hubble parameters, to study cosmic opacity. The
effect of Compton scattering on standard candles is also considered.
The extinction due to Compton scattering can be correctable in
future SNe Ia survey. In order to avoid the cosmological dependence
of SNe Ia luminosity distances, a joint fitting of the SNe Ia
light-curve parameters, cosmological parameters and opacity is used.
In order to explore the cosmic opacity at high redshifts, the latest
gamma-ray bursts (GRBs) are used. Because some instruments will
detect high-redshift GRBs in future, the reionization process must
be considered for Compton scattering. The result shows that the
Compton dimming effect is less than the systematic error for GRBs at
present. However, if more than 100 high-redshift long GRBs are
observed and used to constrain cosmological parameters, the Compton
dimming is non-negligible. The results support an almost transparent
universe at $z < 1.5$ for JLA SNe Ia and $H(z)$ data. In the
redshift range $1.5 < z < 8.1$, we study the cosmic opacity through
luminosity distances of GRBs. The flat $\Lambda$CDM model and the
flat XCDM model are considered. We find that the effect of Compton
scattering can cause about 5\% cosmic opacity in both models. The
current observations support an almost transparent universe for both
cosmological models at a large redshift range.

\section*{Acknowledgements}
We thank the anonymous referee for useful comments. This work is
supported by the National Basic Research Program of China (973
Program, grant No. 2014CB845800), the National Natural Science
Foundation of China (grants 11422325 and 11373022), and the
Excellent Youth Foundation of Jiangsu Province (BK20140016).

\begin{figure}
  \centering
  \includegraphics[scale=0.3,angle=0]{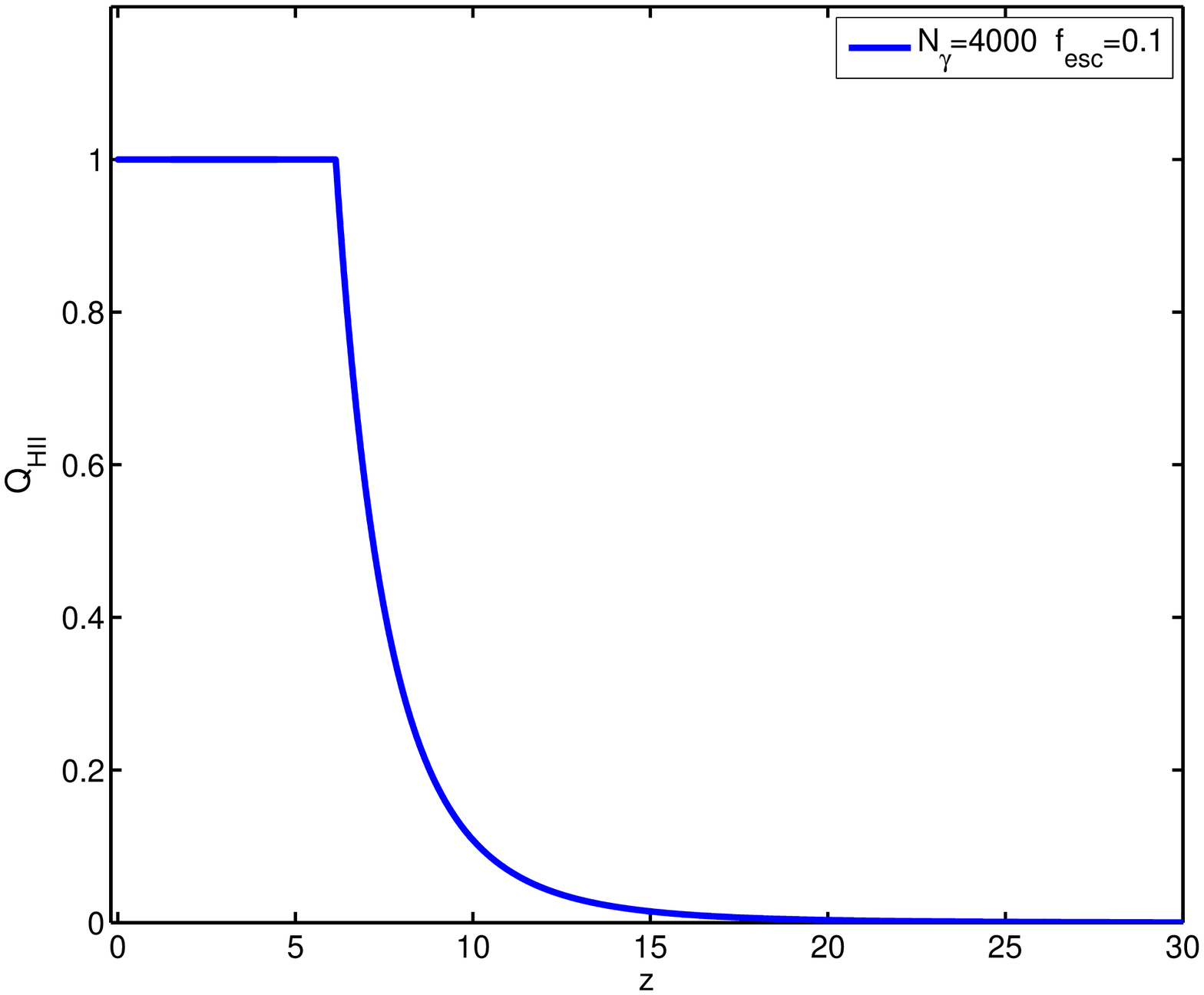}\\
  \caption{$H_{II}$ filling factor $Q_{H_{II}}$ as a function of redshift calculated for the $f_{esc}=0.1$ and $N_\gamma=4000$.}
\end{figure}

\begin{figure}
  \centering
  \includegraphics[scale=0.3]{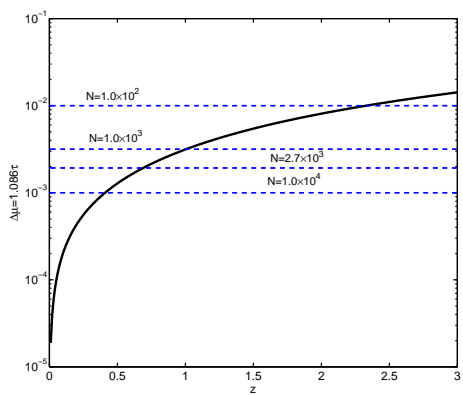}
  \caption{Systematic shift in the distance modulus $\mu$ caused by Compton scattering
(solid line). The dimming is {\rm{ 0}}{\rm{.3\% }} in flux at $z=1$
and $1\%$ at $z=2.35$ with Compton dimming effect.  The statistical
errors for 100, 1000, 2700 and 10000 SNe are shown as the dash
lines. We adopt intrinsic dispersion $\sigma_\mu=0.1$ mag for SNe
Ia.}
\end{figure}

\begin{figure}
  \centering
  \includegraphics[scale=0.5]{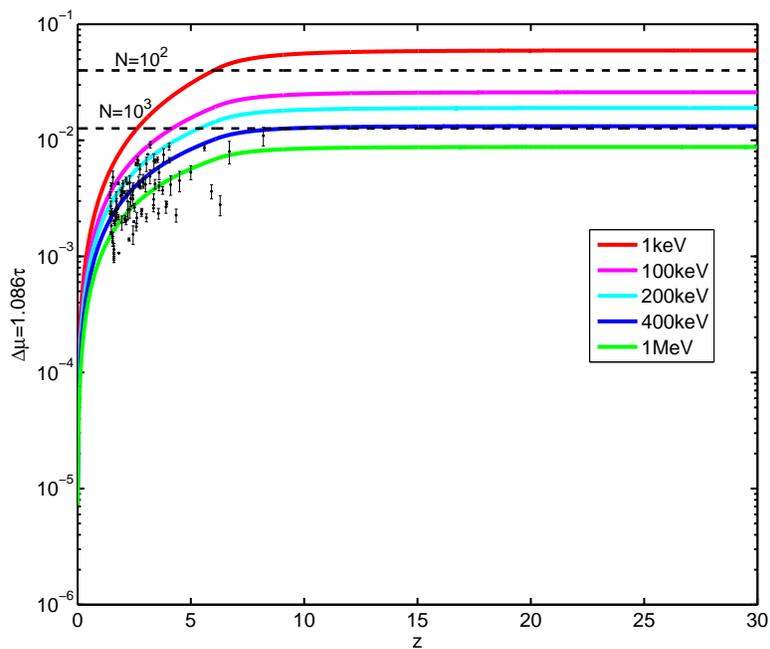}
  \caption{The systematic shift in distance moduli $\Delta\mu$ for GRBs. We consider the ionization
fraction as a function of redshift. The Compton scattering cross
section is energy dependent, because the photons of GRBs are
energetic. The statistical errors for 100 and 1000 GRBs are shown by
the dash lines, respectively. We adopt intrinsic dispersion
$\sigma_\mu=0.4$ mag for GRBs. The black dots are the $\Delta\mu$ of
observed GRBs caused by Compton dimming.}
\end{figure}

\begin{figure}
  \centering
  \includegraphics[scale=0.38]{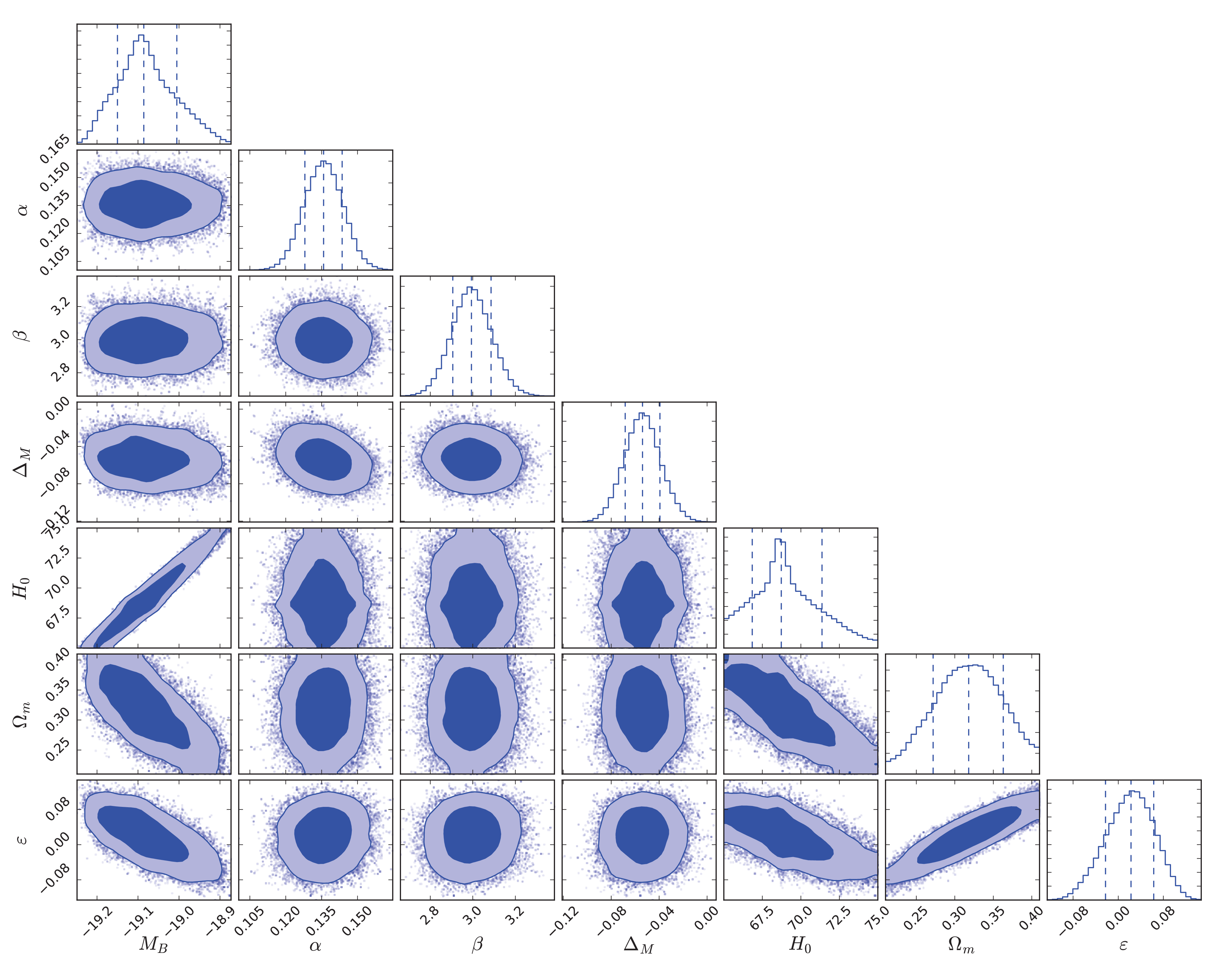}\\
  \caption{In $\Lambda$CDM model, the 2-D regions and 1-D marginalized
  distributions with 1$\sigma$ and 2$\sigma$ contours for the parameters $M_B$,
  $\alpha$, $\beta$, $\Delta_M$, $H_0$, $\Omega_m$, and $\varepsilon$ using SNe Ia+$H(z)$. }
\end{figure}

\begin{figure}
  \centering
  \includegraphics[scale=0.8]{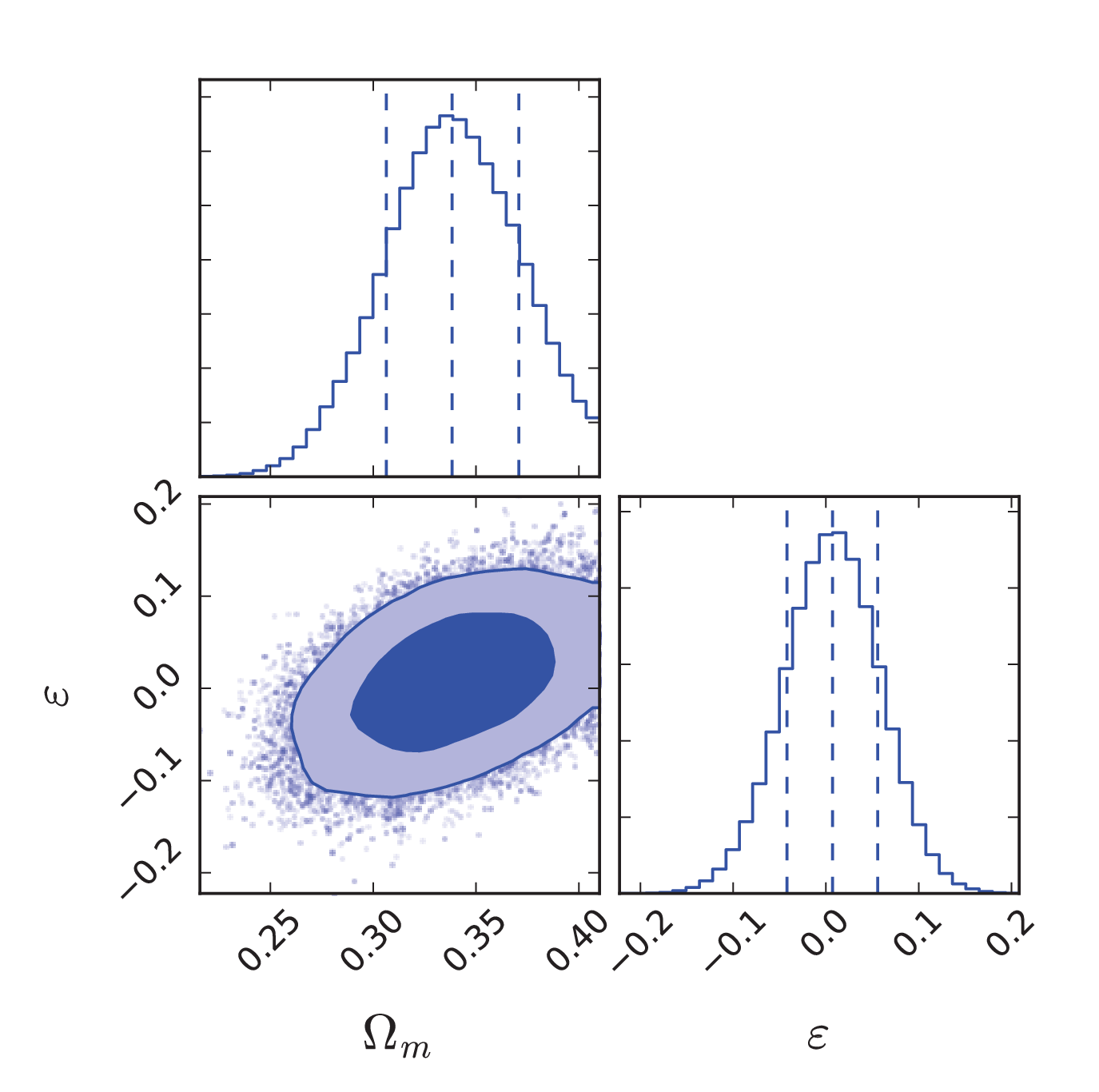}\\
  \caption{The 2 \text{-} D regions and 1 \text{-} D marginalized
  distributions with 1$\sigma$ and 2$\sigma$ contours for the parameters  $\Omega_m$, $\varepsilon$ using GRBs+$H(z)$ in $\Lambda$CDM model. }
\end{figure}

\begin{figure}
  \centering
  \includegraphics[scale=0.38]{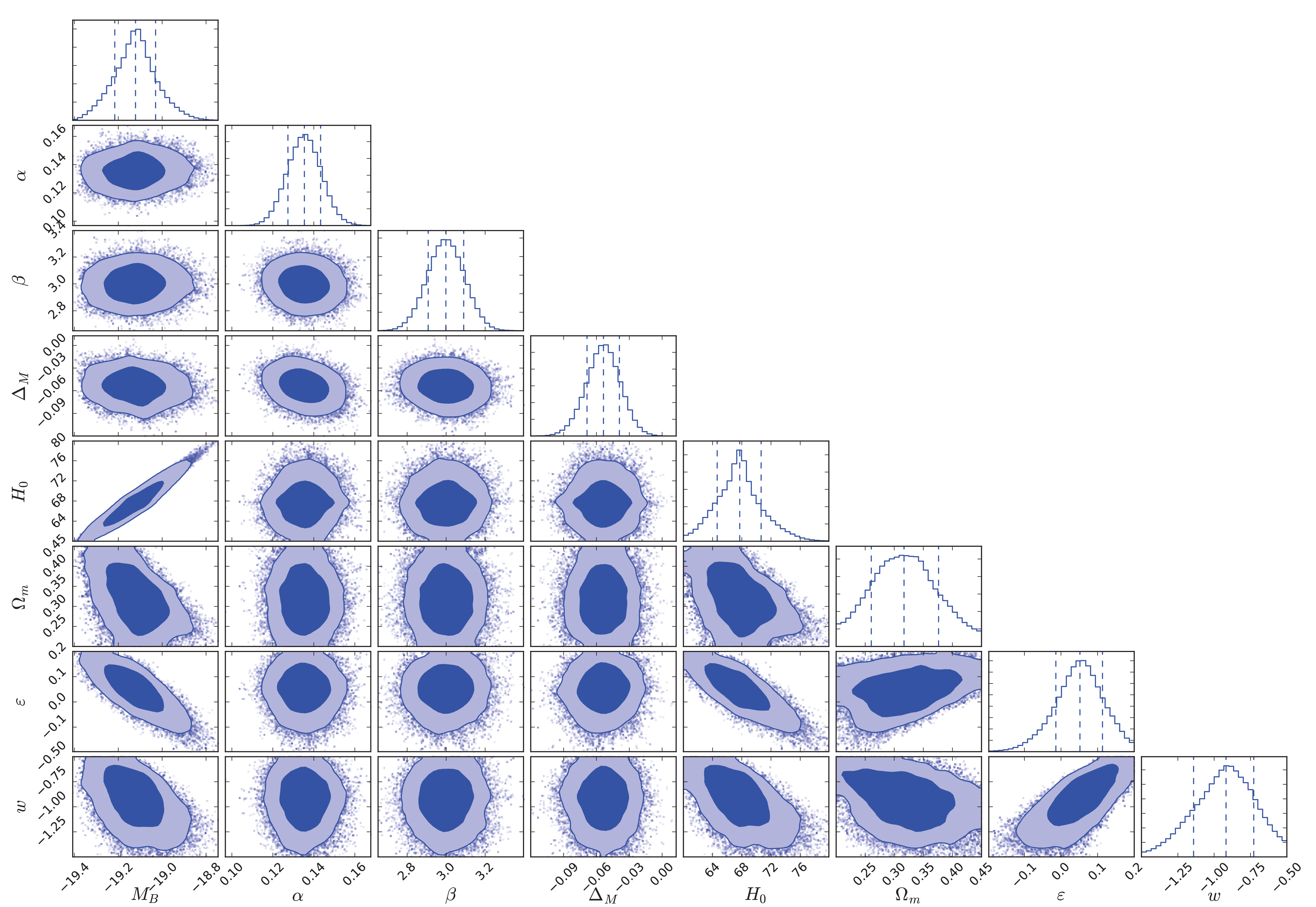}\\
  \caption{In XCDM model, 2 \text{-} D regions and 1 \text{-} D marginalized distributions with 1$\sigma$ and 2$\sigma$
  contours for the parameters $M_B$, $\alpha$, $\beta$, $\Delta_M$, $H_0$, $\Omega_m$, $\varepsilon$ using SNe Ia+$H(z)$. }
\end{figure}

\begin{figure}
  \centering
  \includegraphics[scale=0.7]{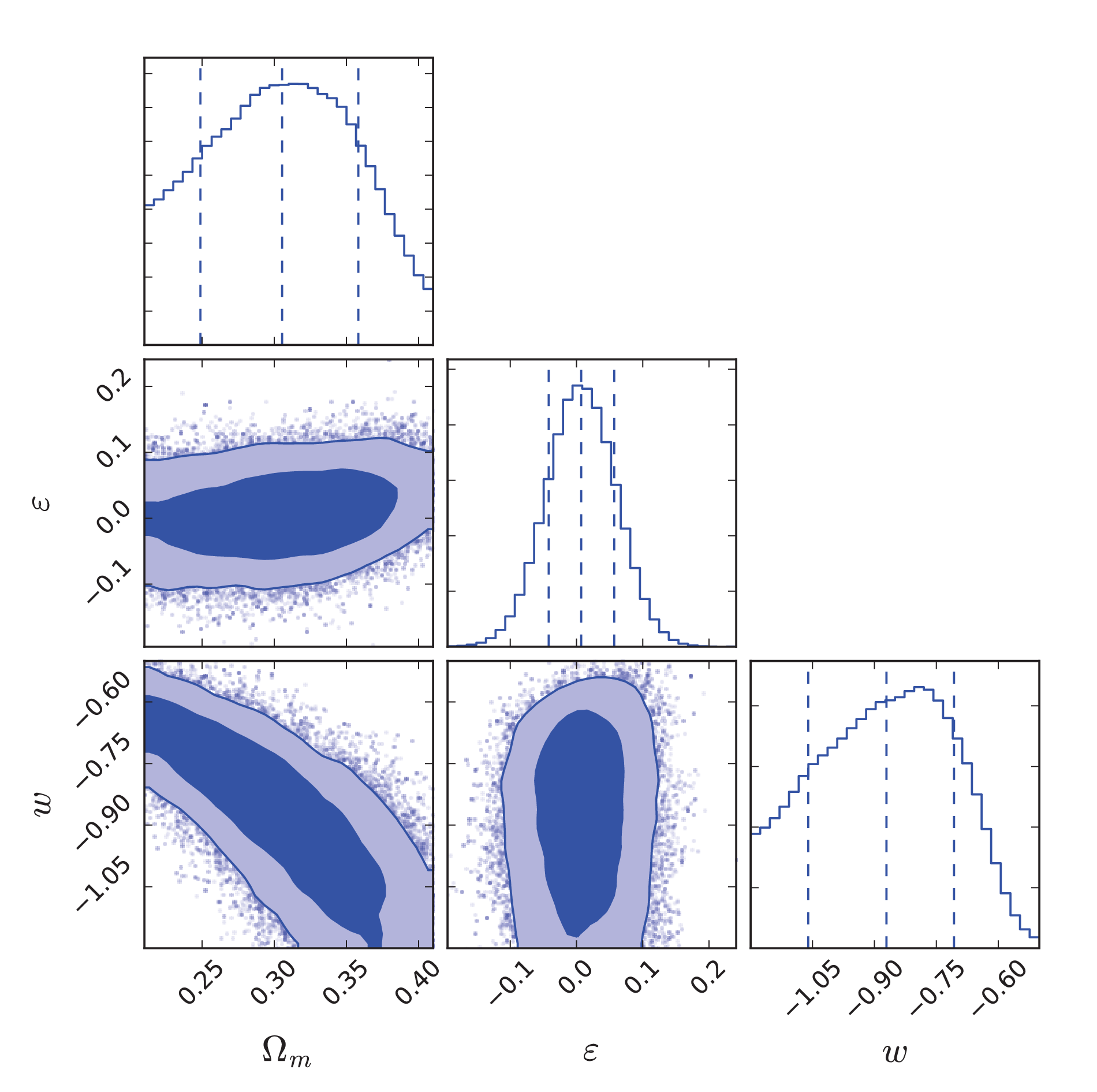}\\
  \caption{In XCDM model, 2 \text{-} D regions and 1 \text{-} D marginalized distributions with 1$\sigma$ and 2$\sigma$ contours
  for the parameters $\Omega_m$, $\varepsilon$ using GRBs+$H(z)$.}
\end{figure}
\begin{figure}
  \centering
  \includegraphics[scale=0.38]{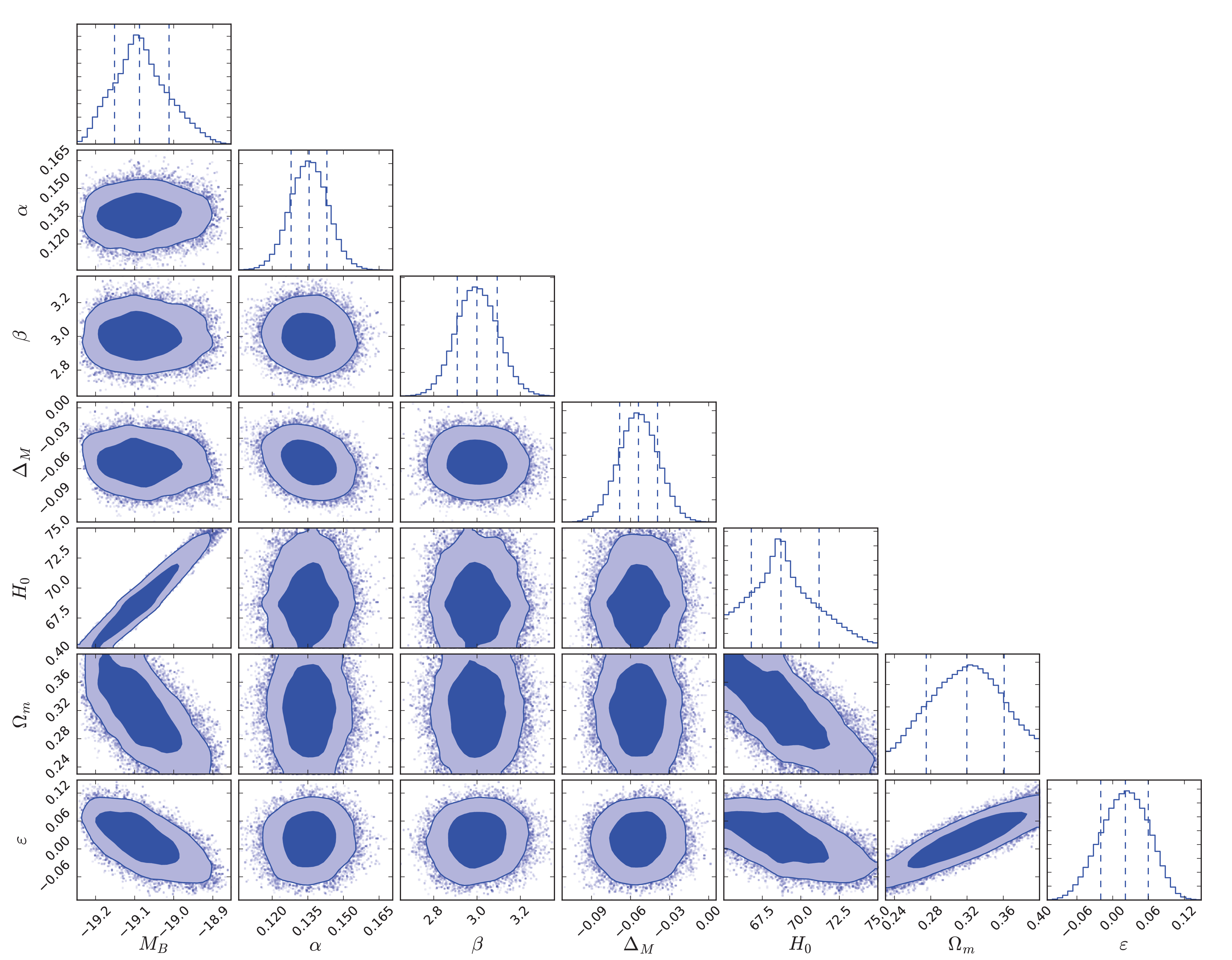}\\
  \caption{In $\Lambda$CDM model, 2 \text{-} D regions and 1 \text{-} D marginalized distributions with 1$\sigma$
  and 2$\sigma$ contours for the parameters $M_B$, $\alpha$, $\beta$, $\Delta_M$, $H_0$, $\Omega_m$, and $\varepsilon$
  using SNe Ia+$H(z)$ after subtracting the effect of Compton scattering. }
\end{figure}

\begin{figure}
  \centering
  \includegraphics[scale=0.38]{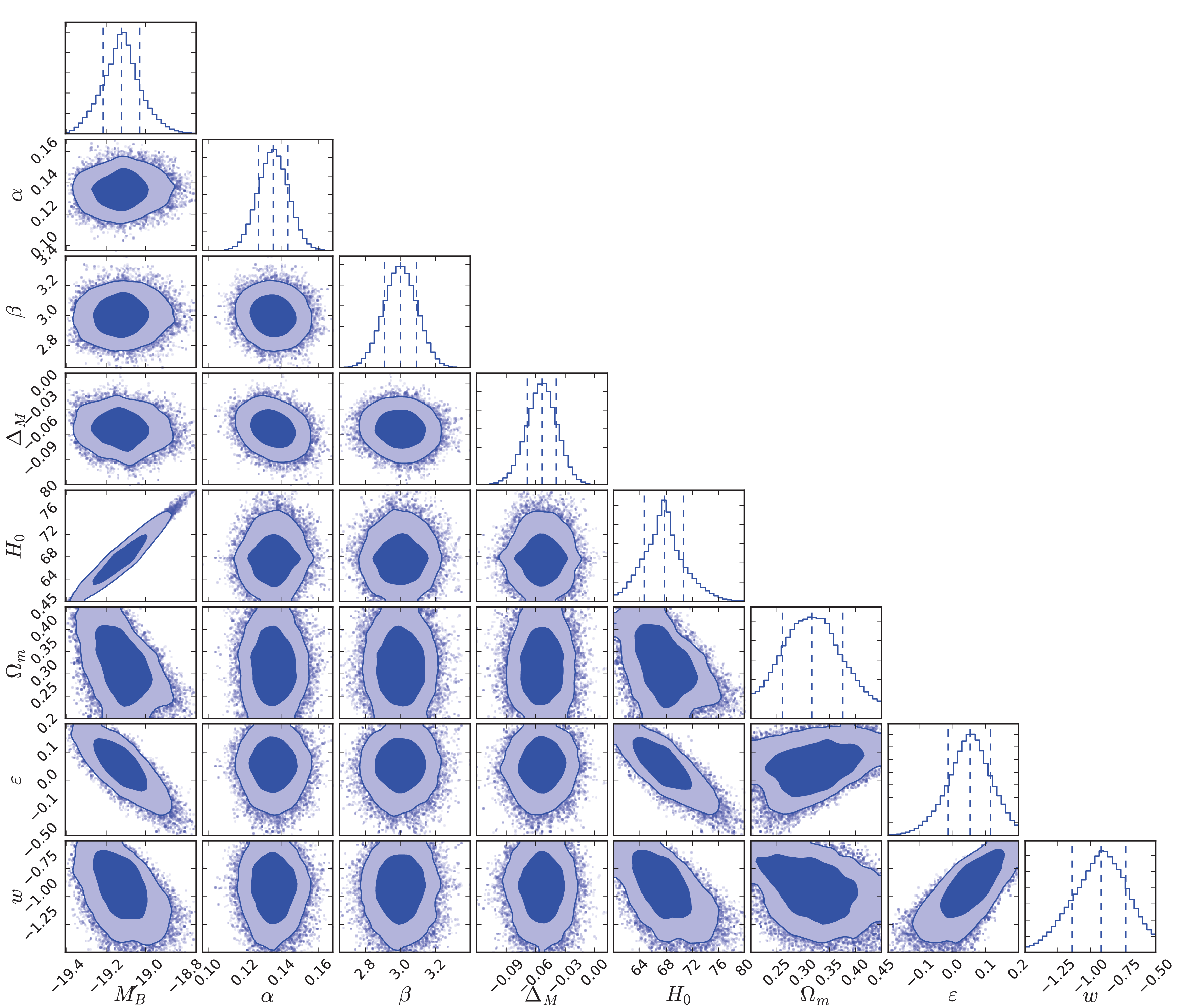}\\
  \caption{In XCDM model, 2 \text{-} D regions and 1 \text{-} D marginalized distributions with 1$\sigma$
  and 2$\sigma$ contours for the parameters $M_B$, $\alpha$, $\beta$, $\Delta_M$, $H_0$, $\Omega_m$, and $\varepsilon$
  using SNe Ia+$H(z)$ after subtracting the effect of Compton scattering. }
\end{figure}

\clearpage

\begin{table*}
%\resizebox{18cm}{!}{}
\caption{Constraints on $\varepsilon$, light-curve parameters and
cosmological parameters with 1$\sigma$ confidence level in different
models.} \scalebox{0.9}{
\begin{tabular} {|c|c|c|c|c|c|c|}
\hline
  Model & \multicolumn{3}{|c|}{$\Lambda$CDM}& \multicolumn{3}{|c|}{XCDM} \\
  \hline
  data set &\multicolumn{2}{|c|}{H(z)+SNe Ia} & H(z)+GRBs & \multicolumn{2}{|c|}{H(z)+SNe Ia}&H(z)+GRBs \\
  \hline
   $\tau(z)$&2$\varepsilon\ln(1+z)$& subtracting $\tau_c$& 2$\varepsilon\ln(1+z)$&2$\varepsilon\ln(1+z)$ & 2subtracting $\tau_c$ & 2$\varepsilon\ln(1+z)$ \\
  \hline
  a&$0.136^{+0.00772}_{-0.00783}$& $0.136^{+0.00749}_{-0.00756}$& $/$ &$0.135^{+0.00790}_{-0.00803}$ & $0.136^{+0.00780}_{-0.00773}$ &$/$\\
  \hline
  b & $2.994^{+0.0923}_{-0.0878}$ & $3.000^{+0.0937}_{-0.0908}$&$/$&$2.999^{+0.0916}_{-0.0907}$&$3.002^{+0.0896}_{-0.0893}$& $/$ \\
  \hline
  $M_B$ & $-19.086^{+0.0807}_{-0.0643}$ & $-19.088^{+0.0756}_{-0.0638}$& $/$ & $-19.121^{+0.0913}_{-0.0948}$ & $-19.113^{+0.0868}_{-0.0854}$ & $/$ \\
  \hline
  $\Delta_M$ & $-0.0539^{+0.0145}_{-0.0145}$& $-0.0539^{+0.0147}_{-0.0144}$& $/$ & $-0.0536^{+0.0146}_{-0.0150}$ & $-0.0543^{+0.0142}_{-0.0152}$ & $/$ \\
  \hline
  $H_0$ & $68.732^{+2.641}_{-1.884}$ & $68.715^{+2.474}_{-1.929}$& $67.8$ fixed & $67.718^{+2.923}_{-3.085}$ & $67.857^{+2.917}_{-2.738}$ & $67.8$ fixed \\
  \hline
  $\Omega_m$ & $0.318^{+0.0449}_{-0.0462}$ & $0.320^{+0.0412}_{-0.0448}$& $0.338^{+0.0325}_{-0.0319}$ & $0.317^{+0.0594}_{-0.0563}$ & $0.308^{+0.0536}_{-0.0553}$& $0.305^{+0.0528}_{-0.0569}$\\
  \hline
  $w$ & $/$ &$/$ & $/$  & $-0.919^{+0.191}_{-0.223}$ & $-0.906^{+0.200}_{-0.205}$ & $-0.871^{+0.164}_{-0.190}$  \\
  \hline
  $\varepsilon$ & $0.0226^{+0.0403}_{-0.0451}$ & $0.0212^{+0.0382}_{-0.0413}$& $0.00718^{+0.0486}_{-0.0492}$ & $0.0517^{+0.0617}_{-0.0659}$ & $0.0490^{+0.0590}_{-0.0654}$ & $0.0718^{+0.0497}_{-0.0491}$ \\
  \hline
\end{tabular}}
\end{table*}

\end{document}